# A Statistical Analysis of Solar Surface Indices Through the Solar Activity Cycles 21-23

Umit Deniz Göker[1], Jagdev Singh[2], Ferhat Nutku[3] and Mutku Priyal[2]

[1]Physics Department, Boğaziçi University, Bebek 34342, Istanbul, Turkey
[2]Indian Institute of Astrophysics, Koramangala, Bengaluru 560034, India
[3]Department of Physics, Faculty of Science, Istanbul University, Vezneciler, Istanbul 34134, Turkey

e-mail: deniz.goker@boun.edu.tr, jsingh@iiap.res.in, fnutku@istanbul.edu.tr and mpriya@iiap.res.in

**Abstract:** Variations of total solar irradiance (TSI), magnetic field, Ca II K-flux, faculae and plage areas due to the number and the type of sunspots/sunspot groups (SGs) are well established by using Solar Irradiance Platform and ground based data from various centers such as Stanford Data (SFO), Kodaikanal data (KKL) and National Geographical Data Center (NGDC) Homepage, respectively. We applied time series analysis for extracting the data over the descending phases of solar activity cycles (SACs) 21, 22 and 23, and the ascending phases 22 and 23 of SACs, and analyzed the selected data using the Python programming language. Our detailed analysis results suggest that there is a stronger correlation between solar surface indices and the changes in the relative portion of the small and large SGs. This somewhat unexpected finding suggest that plage regions decreased in a lower values in spite of the higher number of large SGs in SAC 23 while Ca II K-flux did not decrease by large amount or it was comparable with SAC 22 for some years and relates with C type and DEF type SGs. Thus, increase of facular areas which are influenced by large SGs caused a small percentage decrease in TSI while decrement of plage areas triggered a higher decrease in the flux of magnetic field. Our results thus reveal the potential of such detailed comparison of SG analysis with solar surface indices for understanding and predicting future trends in the SAC.

## 1. Introduction

The variability of solar indices such as total solar irradiance (TSI), the F10.7 cm solar radio flux (F10.7) and the E10.7 ultraviolet flux (E10.7), magnetic field, sunspots, sunspot groups (SGs), facular area (FA), Ca II K-flux and plage area (PA) is very important to understand the structure of the solar activity cycles (SACs).

In general, the magnetic flux emerging in sunspots, faculae or plage and network is investigated by analyzing the effects of dark and bright regions on the solar disc. However, detailed statistics can only be obtained by analyzing the time series of solar surface indices according to group type and sunspot size. As found by the detailed analysis of Kilcik et al. (2011), Lefévre and Clette (2011) and Kilcik et al. (2014), there was an important deficit in small groups after the last activity maximum in 2000 while the number of large groups remained largely stable.

For large SGs (classes D, E, F and G), which contain penumbrae both in the main leading and following spots, the monthly total number of sunspots was similar in both SAC 22 and SAC 23 while the monthly total sunspot area was larger in SAC 22. The average of large sunspot area (SSA) was about 25% smaller in SAC 23 than in SAC 22 (Kilcik et al., 2011). For small SGs types (A and B (no penumbra), C, H and J (at least one main sunspot with penumbra)), the total number of sunspots was higher in SAC 22 than in SAC 23. In addition to this, the monthly total sunspot area was larger in SAC 22 than in SAC 23. The average areas of small sunspots were nearly the same in both SACs. In general, the numbers of large SGs reach a maximum in the middle of the SAC (phases 0.45-0.5), while the international sunspot numbers (ISSN) and the numbers of small SGs generally peak much earlier (SAC phases 0.29-0.35) (Kilcik et al., 2011).

The TSI of the Sun vary mostly because of changes in the areas of dark sunspots and bright faculae (Chapman, Cookson and Dobias, 1997). The number of large and small SGs, and the maximum TSI level for cycle 21 were higher than those for cycle 22 (Kilcik et al., 2011). However, de Toma et al. (2004) mentioned that the TSI did not change significantly from cycle 22 to cycle 23 despite the decrease in sunspot activity of SAC 23 relative to SAC 22.

The TSI, magnetic field, SSA, FA which show better agreement with the number of large SG numbers than they do with the small SG numbers (Kilcik et al., 2011). However, there are two important solar surface indices such as Ca II K-flux and plage regions, which have very important features still need to be analyzed and compared with the variations of large and small SGs. The intensity calibration of Ca II K-line images, however, is very difficult because of the contributions from the quiet Sun limb darkening curve of the Ca II K-line. In addition to this, observations near the core of the ionized Ca II K-line (393.37 nm) are one of the most effective tools to investigate the morphology and evolution of both plages and chromospheric magnetic network (Raju and Singh, 2014). Ermolli et al. (2013) found a linear relation between the PA and sunspot numbers while he found a non-linear relation between the chromospheric plages and photospheric white-light facular areas. Thus, the PA increases while the ratio of faculae area to spot area decreases as the activity level increases.

In the present paper, TSI, magnetic field, sunspots/SGs, Ca II K-flux, FA and PA data are examined, and a time series analysis is done to determine the uncertainty in the last SAC. In this work, different from the literature, we compare the solar surface indices with the types of SGs for the first time and find a relation between different types of SGs and variation of solar surface indices for the SACs. We also define that the explanation of the variability of these indices can be uncompleted without taking any notice of SGs and is studied with the ISSN alone as it has been done before.

The organization of the paper is as follows. In Section 2, we give the data description and explain the data analysis technique. In Section 3, we discuss the data analysis results and in Section 4, we present our conclusions.

## 2. Data description and data analysis technique

The SG classification data were collected mainly from the Rome Observatory for SAC 21 by the United States Air Force/Mount Wilson (USAF/MWL) Observatory database (ftp://ftp.ngdc.noaa.gov/STP/SOLAR_DATA/SUNSPOT_REGIONS/USAF_MWL/docs/sun spotAreaMetadata). All these data are available online at the National Geographical Data Center (NGDC) Website. In addition to this, this database also includes the measurements from the Learmonth, the Holloman and the San Vito Solar Observatories. We used the Learmonth observatory data as the principal source for SACs 22 and 23. The gaps in SG data were filled with records from one of the other stations listed above, so that a nearly continuous time series was produced.

Here, the Zurich classification for cycle 21 and the McIntosh classification (McIntosh, 1990) for cycles 22 and 23 data were used. We followed the similar technique with the work of Kilcik et al. (2011) during the selection of classification but used the different programming language as Python during the analyzes. The data for TSI were taken from the SOLAR2000 data (http://www.spacewx.com/solar2000.html (The Solar Irradiance Platform (SOLAR2000, Research Grade) empirical solar irradiance model is made available, at no charge, to the science and engineering research community) and magnetic field data were taken from Wilcox Solar Observatory (http://wso.stanford.edu/meanfld/MF_timeseries.txt).

The data for faculae and Ca II K-line (393.37 nm) are based on daily synoptic full-disk images obtained by photoelectric photometry of sunspots at the San Fernando Observatory (SFO) (Special contact with Angela M. Cookson). Pixels having a contrast greater or equal to

4.8% were identified as faculae before calculated the correction for foreshortening in millionths of a solar hemisphere (Chapman, Cookson and Dobias, 1997). The plage and Ca II K-flux observations were taken from the Kodaikanal Data (KKL) (Singh et al., 2012; Raju and Singh, 2014; Priyal et al., 2014). However, we preferred using the Ca II K-line images from SFO to coordinate and compare with the faculae observations from the same observatory.

The detection of PA is very hard and only a few observatories have reliable detection of plage data. KKL Observatory is one of these research centers, however, there are some problems with obtaining the data as there are gaps in the data due to bad weather conditions. The long gaps in the data are mainly seen in the rainy season (Priyal et al., 2014). The plage index does not show a clear difference between the northern and southern hemispheres. The northern hemisphere is more active during the ascending years of the cycle but the activity increase in the southern hemisphere during the later years (Raju and Singh, 2014). In spite of the plage observations have some gaps depending on the large amount of difference in weather conditions, KKL observatory has still got one of the good plage data between the stations.

## 3. Data analysis technique

We analyzed all these data and plotted line, scatter and bar plots with the use of Python libraries, Pandas (McKinney, 2010), NumPy & SciPy (van der Walt, Colbert and Varoquaux, 2011) and Matplotlib (Hunter, 2007) in IPython Notebook environment (Perez and Granger, 2007).

Investigated FA (between 1988/4/25-2014/12/29), Ca II K-flux (between 1988/4/25-2014/12/29), PA (between 1986/1/6-2005/12/28), TSI (between 1981/1/1-2008/12/31) and magnetic field (between 1975/5/16-2014/8/3) data are daily, but sunspot counts (between 1982/1/1-2014/5/1) data are monthly averaged. Separately from other indices, we first calculated the absolute value of *magnetic field* because of the opposite polarities and took the average of magnetic field, afterwards. Missing days from time versus magnetic field data were dropped. The analyzing procedure with Python is as follows:

(1) In order to be consistent with the monthly averaged sunspot counts; magnetic field, TSI, FA, Ca II K-flux, PA and the ratio of FA to Ca II K-flux data were grouped for each month by applying the aggregate function of arithmetic mean.

(2) After grouping, sunspot counts and solar surface indices data were joined over the same month and year. The merging operation provide us to compare the data exactly belonging to the same time step.

(3) Cross-correlation analysis between SGs and solar surface indices are applied to unsmoothed merged monthly mean values. Simple moving average of time series was calculated by using **rolling_mean** function of the Pandas. If two data sets were compared over months, data smoothing was done by selecting the window size of the simple moving average as 7 days. This 7-day running mean is clearly seen in the bold continuous curves in all figures throughout the paper. In general, the simple moving average is defined as

$$\bar{x}_i = \frac{1}{2n+1} \sum_{k=i-n}^{i+n} x_k \qquad (1)$$

where *i* is the index used to label an individual observation in the time series, 2*n*+1 is the size of window, in other words, number of observations used in calculating the average. Therefore, taking average of 2*n* points around each $x_i$ and moving through out the time series will give a new time series composed of single moving averages.

(4) Bar plots were generated by grouping the time series data into years and averaging the data of each year by its arithmetic mean. Each type of sunspot counts was normalized by dividing it with the sum of all types of sunspot counts.

(5) After normalization, sunspot counts were grouped by year and averaged for each year. As one can see from bar plots, sum of all normalized averaged sunspot counts for each year is equal to one. For all histograms, the normalization for each year is given by:

$$\frac{AB}{TSSN} + \frac{C}{TSSN} + \frac{DEF}{TSSN} + \frac{H}{TSSN} = 1 \tag{2}$$

For every month, the total sunspot number (TSSN) were calculated, then we divided each sunspot number (SSN) by TSSN. Here "AB" corresponds to A and B type sunspot groups while D, E and F type sunspot groups are mentioned as "DEF" through the paper to make the identifications simple. We preferred to take yearly grouped and average data for plotting bar plots because the monthly averaged data would be too scattered for these kind of plots.

(6) Performing such a normalization allowed us to investigate relative change of the sunspot counts by type for each year. In order to measure whether there is a linear relation between different time series, correlation coefficients were calculated. Pearson product-moment correlation coefficients were calculated by using the Pandas's **corr** function for **dataframes** which is an implementation of the following formula,

$$R_{ij} = \frac{C_{ij}}{\sqrt{C_{ii} * C_{jj}}} \tag{3}$$

where $R_{ij}$ is the correlation coefficient matrix, $C_{ij}$, $C_{ii}$ and $C_{jj}$ are covariance matrices, $i$ and $j$ are the elements of the compared time series. In general, Pearson product-moment correlation coefficient of two independent variable sets containing $n$ items can be calculated by using the following formula,

$$r = r_{xy} = \frac{n\sum x_i y_i - \sum x_i \sum y_i}{\sqrt{n\sum x_i^2 - \left(\sum x_i\right)^2} \sqrt{n\sum y_i^2 - \left(\sum y_i\right)^2}} \tag{4}$$

where $x_1,..., x_n$ and $y_1,..., y_n$ are two independent variable sets and $i$ is the item index. But one should not forget the motto "correlation does not imply causation" while investigating scientific data and drawing a conclusion by using correlation coefficients.

## 4. Discussions and the analysis results
### 4.1. Comparison of sunspot groups with magnetic field and TSI

(4.1.1) To delineate the role of various features of the solar indices, we compare the time variations of these parameters with the different type of SG numbers and SSCs. We further investigate the relationship between these parameters during the decreasing phases of SACs 21-23 and increasing phases of the SACs 22-23. In Fig. 1(a), we plot the sum of smoothed monthly mean number of SGs (SMMNSGs) and monthly mean number of sunspot counts (MMNSSCs) for the sunspot classes "AB, C, H" on the right hand side of the $y$-axis while the left hand side of the $y$-axis corresponds to the scale of magnetic field in the units of $\mu T$. Similarly, Fig. 1(b) shows the variations of the sum of SMMNSGs and MMNSSCs for the sunspot class "DEF" and the total magnetic field with time. In Fig. 2(a), we plot the sum of SMMNSGs and MMNSSCs for the sunspot classes "AB, C, H" on the right hand side of the $y$-axis while the left hand side of the $y$-axis corresponds to TSI in the units of $W/m^2$ and in Fig.

2(b), it shows the variations of the sum of SMMNSGs and MMNSSCs for the sunspot class "DEF" and TSI as a function of time.

(4.1.2) In Fig. 1, the number of AB (simple group of sunspots) SGs in the maximum phase of SAC 22 is higher than in SAC 23 (SAC 22 > SAC 23) and it is the same in the minimum phase (SAC 22 > SAC 21 > SAC 23) and these SGs did not increase even in the beginning of SAC 24. The number of C (medium group of sunspots) type SGs is also higher in the maximum of SAC 22 than in SAC 23 (SAC 22 > SAC 23) and in minimum it decreased to zero values for some years in SAC 23 as it is similarly seen in H groups. In DEF groups, there has been no important difference between all cycles and the number of these groups in the first and second maximums are SAC 22 $\gtrsim$ SAC 23 and SAC 22 $\lesssim$ SAC 23 respectively while the minimum in SAC 21 and SAC 22 had similar decreasing than as in SAC 23. Thus, we can define that the affect of the large SGs was more dominant in the last SAC. H groups showed similar numbers in the maximum phases of SAC 22 and SAC 23 (SC 22 $\gtrsim$ SC 23) and no distinct data in the decreasing phase for some observing days.

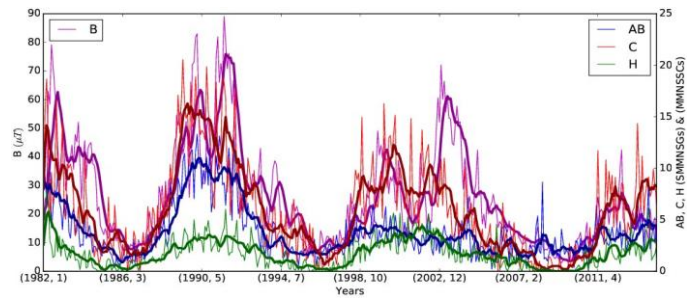

(a) Time series of AB, C and H type SGs and SSCs are compared with absolute value of the monthly grouped and averaged magnetic field. Thick lines are 15 day simple moving averages.

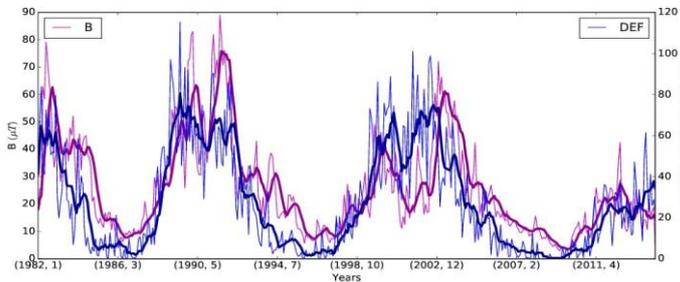

(b) Time series of DEF type SGs and SSCs are compared with absolute value of the monthly grouped and averaged magnetic field. Thick lines are 15 day simple moving averages.

**Figure 1:** The *x*-direction denotes the years and the *y*-direction denotes the monthly mean SGs and SSC numbers (right) and monthly mean magnetic field (left). The magnetic data start from January, 16 1982 and complete May, 1 2014 while SGs and SSC numbers start from January, 16 1982 to May, 1 2014. The SG and SSC numbers of AB, C and H in Fig. (a) and DEF in Fig. (b) are shown with blue, red, green and violet, respectively while magnetic field is shown with purple color in both figures.

(4.1.3) The TSI profile for cycle 23 has a distinct double peak, and the second peak began in the middle of 2001 and it approximately finished at the end of 2002, thus coinciding with the peak of the large SG numbers while the double peak in SAC 22 showed comparable increasing. In SAC 23, large SGs reach their maximum numbers about 2 years later than the small ones as seen in Figs. 1 and 2. This time difference is more effective for the magnetic field than in TSI. This time delay proves at least, partially, due to an excess of large SGs present on the solar disk during the second half of cycle 23. Small groups are efficient in TSI after the maximum in SAC 22 (in the beginning of 1990) and in the maximum of SAC 23 (in the year 2000) as seen in Fig. 2(a) while large SGs show double peak both in SAC 22 and SAC 23 as shown in Fig. 2(b). However, large SGs were dominant after the year 2002 in SAC

23 and this situation influenced in the FA and sunspot area. The question here is that why was the TSI value still comparable with the SAC 22 while the large SGs were higher?

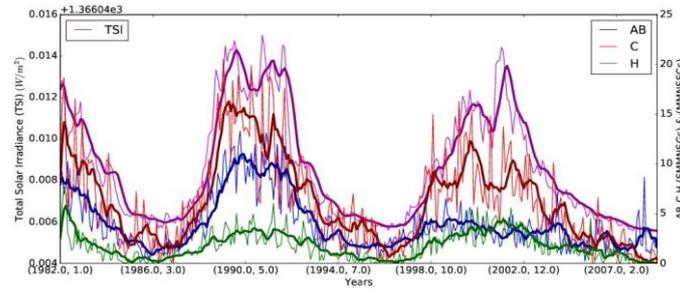

(a) Time series of monthly grouped and averaged TSI are compared with AB, C and H type SGs and SSCs. Thick lines are 15 day simple moving averages.

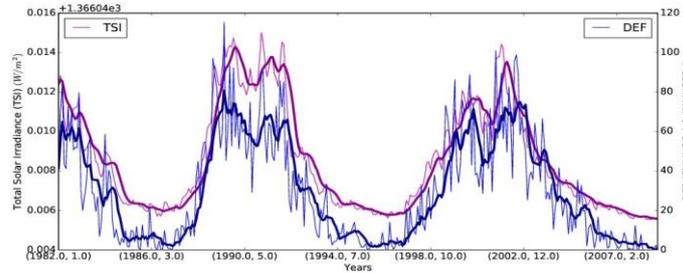

(b) Time series of monthly grouped and averaged TSI are compared with DEF type SGs and SSCs. Thick lines are 15 day simple moving averages.

**Figure 2:** The *x*-direction denotes the years and the *y*-direction denotes the monthly mean SGs and SSC numbers (right) and monthly mean TSI (left). The TSI start from January, 16 1982 and complete December, 16 2008 while SGs and SSC numbers start from January, 16 1982 to December, 16 2008. The SGs and SSC numbers of AB, C and H in Fig. (a) and DEF in Fig. (b) are shown with blue, red, green and violet, respectively while TSI value is shown with purple color in both figures.

(4.1.4) In addition to the question above, the magnetic field was distinctively lower in SAC 23. Both for small and large SGs started to appear before the magnetic field increasing occurs in SAC 23 while SGs and magnetic field change in a synchronized way in SAC 22 as seen in Figs. 1(a) and 1(b). In Fig. 1(a), the maximum for SAC 22 is seen at the end of 1991 and C type SGs are correlated with the first peak of the magnetic field in SAC 22. However, there is a huge difference in the second peak of the solar maximum 23 (in the middle of 2003) with C type sunspot groups. All small type SGs were lower in the SAC 23 as compared to large SGs. In Fig. 1(b), we can easily see that the first (at the end of 1989) and the second (at the end of 1991) peaks are correlated with large SGs, in contrast to this, magnetic field value is substantially lower in SAC 23 while large SGs are still comparable with SAC 22. Another question here is that why did the magnetic field decrease in that situation?

(4.1.5) Lefévre and Clette (2011) found that there were fewer short-lived groups of types A, B and a part of C in SAC 23 than in SAC 22. A and B types were largest in the lower lifetime ranges (<5 days). Only C type group lifetimes had a wide and rather flat distribution intermediate between small A and B groups and large groups, suggesting a dual population. Lefévre and Clette (2011) also found that there were no significant difference in the distributions for D to F groups, even in the short lifetime tails and our analysis results prove all these subtractions. So, we applied the same analyzing technique for FA and PA, and Ca II K-flux which are the main indices will shed light to our understanding of SACs. All the answers of the questions above will be correlated with the analyzes of these parameters.

## 4.2. Facular Area and Ca II K-flux

(4.2.1) In Figs. 3(a) and 3(b), we plot the sum of SMMNSGs and MMNSSCs for the sunspot classes "AB, C, H" and "DEF" on the right hand sides of the *y*-axis respectively while the left hand sides of the *y*-axis correspond to the values of Ca II K-flux. Similarly, in Figs. 3(c) and 3(d), we plot the sum of SMMNSGs and MMNSSCs for the sunspot classes "AB, C, H" and "DEF" on the right hand sides of the *y*-axis respectively while the left hand sides of the *y*-axis correspond to the variations of FA. The Ca II K-flux was obviously lower in SAC 23 both for small and large SGs. In addition to this, Ca II K-flux variations are comparable with C and DEF type SGs, however, large SGs show higher correlation than the small SGs as given in Figs. 3(a) and 3(b). Large SGs also have a correlation with the peaks in the maximum of SAC 22 and SAC 23 as seen in Fig. 3(b).

(4.2.2) In Fig. 3(a), the small SGs and Ca II K-flux variations change synchronized in SAC 22 and SAC 23, however, the Ca II K-flux variation is 30% lower in SAC 23 than in SAC 22. Small AB groups correlated with Ca II K-flux in SAC 22 while they moved away from the correlation in SAC 23 (in the year 1999). The first peak is not clearly seen for Ca II K-flux in the last solar cycle (in the year 2000) and second peak starts in the beginning of the year 2003 as seen in Fig. 3(b). FA and Ca II K-flux were separated from synchronization in the beginning of 1999. FA started to increase since the beginning of 1999 until the middle of 2009 and it started to increase again after the year 2010.

(4.2.3) The double peak in the last SAC (SAC 23) for FA was distinctively clear than for Ca II K-flux and the level of the peak was higher, as well. The small SGs (especially C groups) were comparable with the variation of Ca II K-flux while we did not see such correlation for the FA. The number of small C type SGs and their coverage on the solar surface were smaller while the total area of faculae covered more surface on the Sun. The second important thing is that the Ca II K-flux was lower in SAC 23 than in SAC 22 while the FA were comparable both for SACs 22 and 23.

(4.2.4) In Fig. 3, the coverage of FA on the solar surface is maximum during the year 1990 in SAC 22 while it is lower for SAC 23 in the year 2001. Similarly, the same decreasing was seen for Ca II K-flux, namely the maximum coverage for SAC 22 in 1991 was higher than for SAC 23 in 2001 and it continuous along with the decreasing phase of SAC 23. The decrease of both FA and the number of small SGs in the weaker cycle 23 may have been compensated by the higher number of large SGs. FA highly correlate with the large SGs for both SACs rather than with the small groups as seen in Fig. 3(d). The maximum peaks in the years 1990-1992 for SAC 22 and 2001-2003 for SAC 23 correlated with the large SGs, as well. This shows that large SGs follow SACs well-ordered than the small SGs.

(4.2.5) The area ratio of faculae to Ca II K-flux was 60% in the SAC 22 and 69% in the SAC 23, this is because the decrease of Ca II K-flux in the last solar cycle while FA did not show significant difference. The coverage of FA on the solar surface decreased only 20% from SAC 22 to SAC 23, however, this decreasing was 30% for Ca II K-flux and this caused approximately 70% difference in total between the two important SACs.

(4.2.6) As for the bar plots in Fig. (4), we analyze all these yearly variations we mentioned above with their percentage variations in one column and all boxes here give the percentage share of the parameters. The peak of large SGs count delay relative to the solar cycle maximum. FA are declining with time and the decrease in peak cycle amplitude is also seen in the normalized average of sunspot counts. During the years 2002-2008, DEF ratio is decreasing with decreasing FA, however, AB, C and H ratio is increasing. The FA ratio decreases with increasing activity levels. The PA correlates with the C type of SGs stronger than the other types.

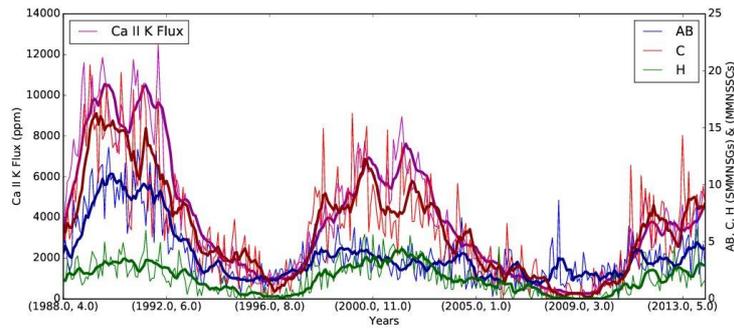

(a) Time series of monthly grouped and averaged Ca II K-flux are compared with time series of AB, C and H type SGs and SSCs. Thick lines are 15 day simple moving averages.

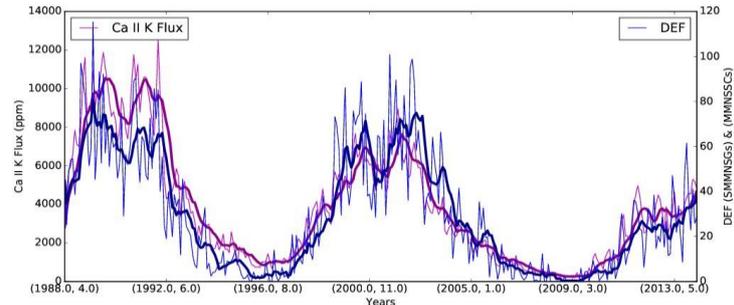

(b) Time series of monthly grouped and averaged Ca II K-flux are compared with time series of DEF type SGs and SSCs. Thick lines are 15 day simple moving averages.

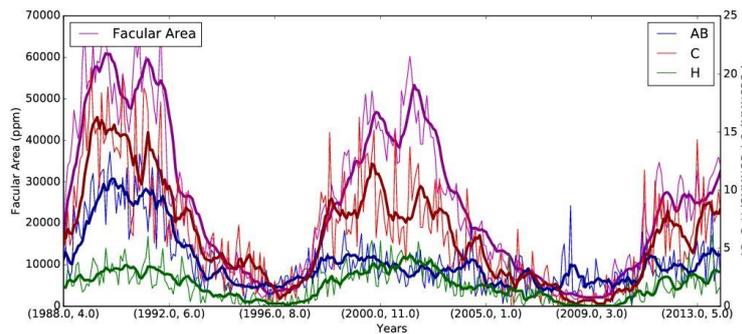

(c) Time series of monthly grouped and averaged FA are compared with time series of AB, C and H type SGs and SSCs. Thick lines are 15 day simple moving averages.

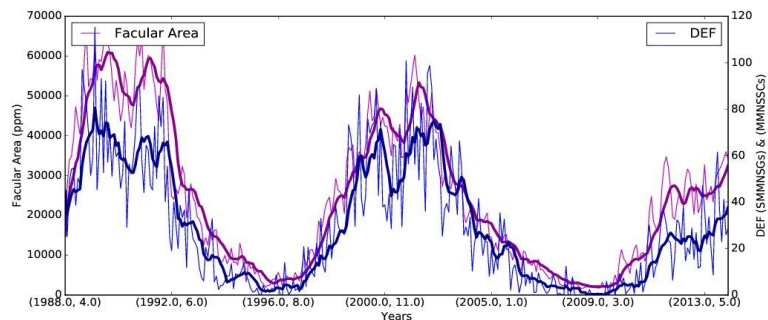

(d) Time series of monthly grouped and averaged FA are compared with time series of DEF type SGs and SSCs. Thick lines are 15 day simple moving averages.

**Figure 3:** The *x*-direction denotes the years and the *y*-direction denotes the monthly mean SGs and SSC numbers (right) and Ca II K-flux (a, b) and FA (c, d) (left). The Ca II K-flux and FA data start from April, 25 1988 and complete April, 3 2014 while SGs and SSC numbers start from April, 25 1988 to April, 3 2014. The SGs and SSC numbers of AB, C, H and DEF in Figs. (a-d) are shown with blue, red, green and violet, respectively while Ca II K-flux and FA are shown with purple color in both figures.

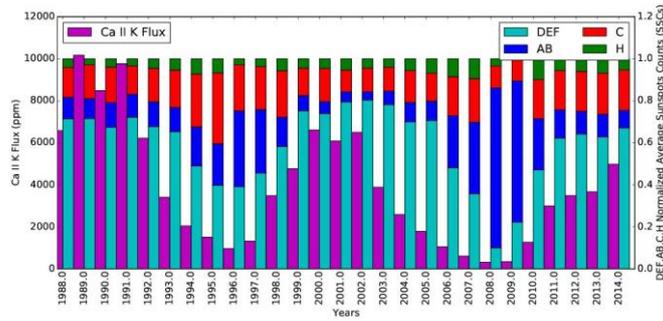

(a) Comparison of Ca II K-flux with normalized average sunspots counts in a time scale of a year is shown. The time intervals correspond to the Ca II K-flux are June, 15 1988-March, 1 2014.

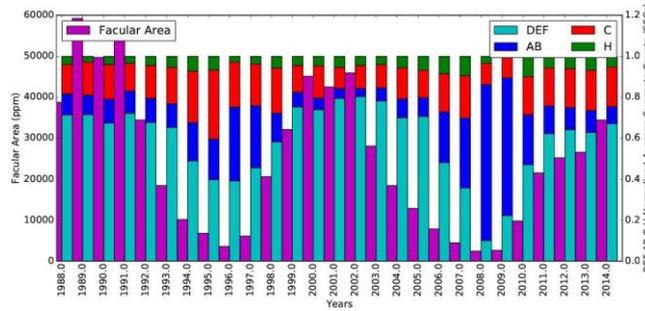

(b) Comparison of FA with normalized average sunspots counts in a time scale of a year is shown. The time intervals correspond to the FA are June, 15 1988-March, 1 2014.

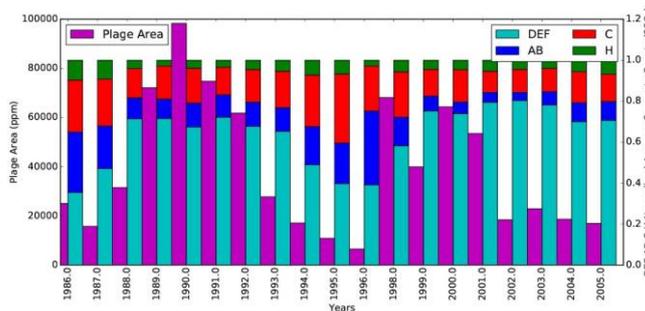

(c) Comparison of PA with normalized average sunspots counts in a time scale of a year is shown. The time intervals correspond to the PA are June, 15 1986-June, 15 2005.

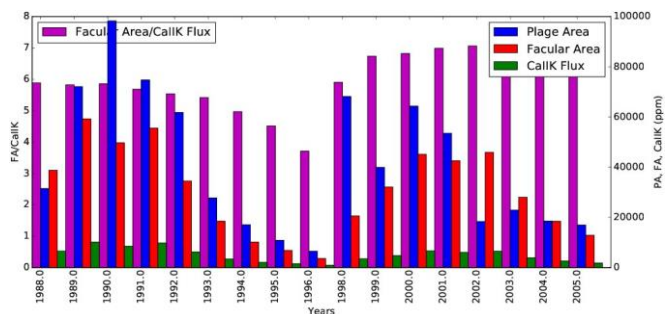

(d) Comparison of grouped ratio of FA to Ca II K-flux with PA and FA, and Ca II K-flux in a time scale of a year is shown. The horizontal axis corresponds to the period August, 3 1988-June, 27 2005.

**Figure 4:** Variations of monthly mean Ca II K-flux, FA and PA data and sunspot numbers versus years are given. The horizontal axis corresponds to the period June, 15 1988-March, 1 2014 for the Figs. 4(a) and 4(b) while it corresponds to the period June, 15 1986-June, 15 2005 and August, 3 1988-June, 27 2005 for the Figs 4(c) and 4(d), respectively. The *y*-axis on the left-hand sides show the Ca II K-flux, FA and PA with the purple color while the averaged SSC numbers for AB, C, H and DEF classes are indicated on the right-hand sides of the *y*-axis with blue, red, green and light-green colors for the Figs. (a), (b) and (c), respectively. In addition to this, the *y*-axis on the left-hand side shows the grouped ratio of FA to Ca II K-flux with the purple color while the PA and FA, and Ca II K-flux are the blue, red and green colors are indicated on the right-hand side of the *y*-axis in Fig. (d).

(4.2.7) In Figs. 4(a), 4(b) and 4(c), we plot the normalized average sunspot counts (SSCs) for the sunspot classes "AB, C, H" and "DEF" on the right hand sides of the *y*-axis while the left hand sides of the *y*-axis correspond to the values of Ca II K-flux, FA and PA, respectively. In the bar plot 1 (Fig. 4(a)), the decreasing is clear for Ca II K-flux in the SAC 23 and the SAC 24 also has a lower values than SAC 23 during the ascending phase. The highest Ca II K-flux value was seen for SAC 22. This histogram shows that in the descending phase of SAC 22, the Ca II K-flux decreases with the decreasing percentage of SGs as C > DEF > AB from 1991 to 1995, however, the total number of large SGs are still higher; in the increasing phase of SAC 23, the Ca II K-flux is increasing with DEF SGs, the highest number of increasing is seen for AB SGs in the first two years but large SGs keep the increasing in their hands from 1996 to 2000. The yearly percentage of the SGs are shown as DEF > AB > C in the decreasing phase of SAC 23, the Ca II K-flux highly decreased and still does not climb to the highest levels. From the year 2001 to 2005 the distribution of SGs were DEF > C > AB, however, AB type SGs started to increase their number from 2006 to 2009 while the percentage of large SGs were decreased. The maximum number of AB SGs were seen in the years 2006-2009, afterwards they decreased in an abrupt manner from the year 2010.

(4.2.8) In Fig. 4(b), FA and Ca II K-flux show similar variations in the increasing phase of SAC 22 from the year 1988 to 1990; in the decreasing phase of SAC 22, the FA is higher in the year 1991 but it is still comparable for either Ca II K-flux and the number of SGs; in the increasing phase of SAC 23, FA started to increase from 1996 to 2000. When we compare the total variation of Ca II K-flux and FA, the FA did not change significantly both in SACs 22, 23 and the beginning of SAC 24, and FA correlates with large SGs very well. The total area of faculae was higher even in the decreasing phase of SAC 23 (from the beginning of 2001) opposite to Ca II K-flux. The average area of Ca II K-flux (and faculae) in the descending phase of SAC 22, increasing phase of SAC 23, decreasing phase of SAC 23 and the beginning of the increasing phase of SAC 24 are 37% (63%), 45% (56%), 13% (87%) and 13% (87%), respectively when we compared Figs. 4(a) and 4(b).

(4.2.9) In the bar plot 4 (Fig. 4(d)), we plot the grouped ratio of FA to Ca II K-flux on the left hand side of the *y*-axis while the right hand side of the *y*-axis corresponds to PA, FA and Ca II K-flux in a time scale for the given years in the *x*-axis. As clearly seen from the Fig. 4(d), the *total* area ratio of faculae to Ca II K-flux, the total area of plages, faculae and Ca II K-flux percentages are given as 31%, 47.44%, 42% and 34.43% and 46%, 31.3%, 43.87% and 53.28% in the descending phase of SAC 22 and SAC 23, respectively. Thus, we can see that only the total area of plages were decreased in SAC 23 (SAC 22 > SAC 23), in contrast to this, SAC 22 < SAC 23 for all other parameters (especially for Ca II K-flux). As for the *average* values of the area ratio of faculae to Ca II K-flux, the area of plages, faculae and Ca II K-flux percentages are given as 31.8%, 47.40%, 43.66% and 36.78% and 39%, 26%, 38% and 47.46% in the descending phases of SACs 22 and 23, respectively. When we compared the descending phases of SAC 22 and SAC 23, the highest values were found for the average area of plages in SAC 22 and only the average area of faculae was decreased in SAC 23 in contrast to the total area of faculae.

It can be easily seen that the visibility of the total area of faculae and Ca II K-flux was very low in the ascending phase of SAC 23 and the average area of faculae and Ca II K-flux had similar structure, as well. Therefore, it can be mentioned that the covered area of faculae and Ca II K-flux on the solar surface was small, however, the PA was still influence in the ascending phase of SAC 23.

(4.2.10) We let explaining the third bar plot to the end because we wanted to compare these results with the following section which describes the plage data variations. As clearly seen from the Fig. 4(c), the total area of plages decrease both for the descending and ascending phases of SAC 23. The most important decrease was seen in the descending phase of SAC 23 (SAC 22 (38%) » SAC 23 (25.3%)), however, it was comparable in the ascending phase of SAC 23 (SAC 22 (19.6%) ≈ SAC 23 (17%)). For the average PA on the solar surface, SC 22 was dominant than the last SAC, it was distinctively decreased in the descending phase of SAC 23 (SAC 22 (36%) » SAC 23 (20%)) while it was comparable in the ascending phase of SAC 23 (SAC 22 (23.4%) ≈ SAC 23 (20.4%)).

When we compare the bar plots of PA with the averaged sunspot counts, they are mostly related with DEF and C type SGs rather than AB and H SGs. Especially, in the descending phases of SAC 22 and SAC 23, DEF SGs reach their maximum values and C type SGs follow these DEF SGs in the second order. However, AB type SGs are in the lowest values for the descending phase of SAC 23 than the other phases of SAC 22 and SAC 23, and they are in the same number in the ascending phases of SAC 22 and SAC 23. We must remember one important thing that there were no available plage data for the whole year of 1997, so, we did not have any chance to compare all the other parameters we mentioned above with the plage data for the year 1997 for the bar plots in Figs. 4(c) and 4(d). We have to ask one important question here that why do the PA in a low numbers while DEF SGs are higher numbers in all SACs? What did cause this decreasing in the PA?

### 4.3. Plage regions

(4.3.1) In Fig. 5(a), we plot the sum of SMMNSGs and MMNSSCs for the sunspot classes "AB, C, H" and "DEF" on the right hand side of the *y*-axis while the left hand side of the *y*-axis corresponds to the values of the monthly grouped and averaged PA. In Fig. 5(a), it is clearly seen that DEF SGs start their rising before the PA evaluate, however, the other SGs follow the evolution steps of PA. In addition to this, the covered area of plages on the Sun decrease after the middle of the year 2002 while DEF SGs are still in a higher values. This is the first important result for the comparison of PA with SGs from our data analysis. This figure shows that the connection between DEF SGs and PA is in a lower values (in spite of the high correlation of solar surface indices with DEF SGs) than the connection between DEF SGs and FA.

(4.3.2) The other important result we find from the Fig. 5(a) is that large SGs have peaks in the maximum phase of SACs 22 and 23 while small SGs are following regular variation. For example, we can see the first peak in the increasing phase of SAC 22 (in the beginning of 1990) and the second peak in the decreasing phase of SAC 22 (in the beginning of 1992) for DEF SGs while the first and only peak is seen for the PA in the middle of 1991. Another example can be given in the descending phase of SAC 23, the first peak is seen in the end of 2000 and the second peak is seen in the middle of 2002, however, the peaks for PA are closer to each others and they are seen approximately in the beginning of 2001. This shows that PA have different structure than the other parameters as FA and Ca II K-flux. So, we must compare all these indices together if we want to understand SACs.

(4.3.3) In Fig. 5(b), we plot the monthly grouped and averaged FA and Ca II K-flux on the right hand side of the *y*-axis while the left hand side of the *y*-axis corresponds to the values of the monthly grouped and averaged PA. In Fig. 5(b), we compare the variation of FA, Ca II K-flux and PA in the ascending and descending phases of SAC 22 and SAC 23. From this analysis, we find that FA reach the first peak in the middle of 1990 and second peak in the beginning of 1992 for the increasing phase of SAC 22, however, plage regions show only one peak in the beginning of 1991. The FA show first peak in the beginning of 2001 and second peak in the middle of 2002, however, PA show very close (double) peaks in the middle of 2001 for the SAC 23. The FA evaluated before the evolution of plage regions, so they would reach maximum phases before than the PA. This figure again shows that FA correlates with large SGs (even in the peaks) in a highly numbers rather than the correlation between PA and large SGs. Thus, we can say that we must analyze the variations in SACs for faculae, Ca II K-flux and plage regions together.

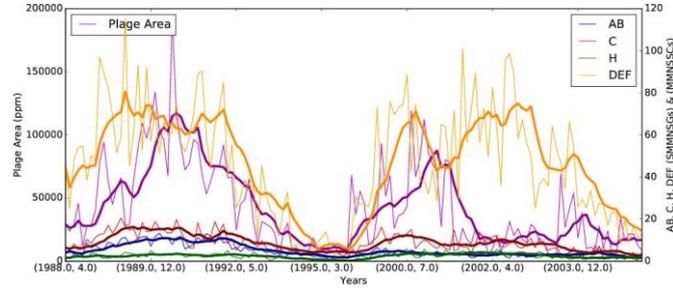

(a) Time series of monthly grouped and averaged PA compared with time series of AB, C, H and DEF type sunspot counts are given. Thick lines are 15 day simple moving averages.

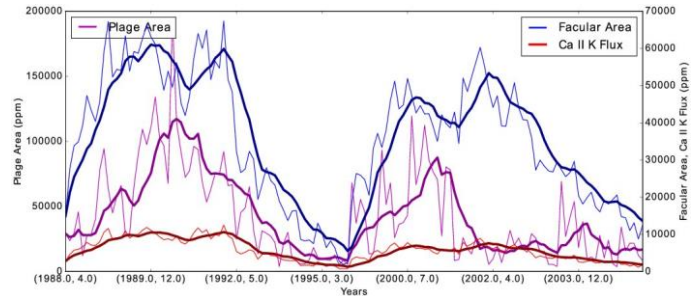

(b) Time series of monthly grouped and averaged PA compared with monthly grouped and averaged FA and Ca II K-flux are given. Thick lines are 15 day simple moving averages.

**Figure 5:** The *x*-direction denotes the years and the *y*-direction denotes the monthly mean SGs and SSC numbers (right) and the monthly grouped and averaged PA (left) in Fig. (a). Besides, the *y*-direction denotes the monthly grouped and averaged FA and Ca II K-flux (right) and the monthly grouped and averaged PA (left) in Fig. (b). The SSCs data start from April, 17 1988 and finish in December, 20 2005 while PA and FA, and Ca II K-flux data start from April, 17 1988 till to December, 20 2005. In Fig. 5 (a), the PA is shown with purple color while AB, C, H and DEF are shown with blue, red, green and yellow colors, respectively. In Fig. 5 (b), the PA is indicated with purple color while FA and Ca II K-flux are shown with blue and red color, respectively.

In order to summarize correlations, we made a table (Table 1) which includes Pearson correlation coefficients between monthly averaged SGs, and monthly grouped and averaged magnetic field (absolute value of it), TSI, FA, Ca II K-flux and PA. As seen in Table 1, the Pearson correlation coefficients of TSI, FA and Ca II K-flux correlates with the type of SGs with their high number of correlation values. However, |B| and PA coefficients have lower correlation values. Especially in PA, these lower values can easily be seen for DEF and H type SGs. Thus, this table shows that the lower number of correlations for some indices are not caused with the number and the size of SGs alone, there are additional effects we must consider for this variations. In addition to this, if this decreasing only related with the SGs, the correlation results would be the same for TSI, FA and Ca II K-flux, as well.

**Table 1:** Pearson Correlation Coefficients between Solar Indices (SI) and AB, C, H and DEF type SGs.

| SI | AB | C | H | DEF |
|---|---|---|---|---|
| $|B|$ | 0.626 | 0.632 | 0.455 | 0.630 |
| TSI | 0.736 | 0.840 | 0.754 | 0.896 |
| FA | 0.708 | 0.859 | 0.752 | 0.890 |
| Ca II K-Flux | 0.760 | 0.864 | 0.704 | 0.869 |
| PA | 0.612 | 0.595 | 0.266 | 0.411 |

The lower correlation coefficients both for |B| and PA prove our results because as we know the magnetic field strongy relates with the plage regions. Our data was limited for plage regions depending on the weather conditions, on the other hand, we have full data for the magnetic field. Accordingly, this lower correlation coefficients do not relate with the number of data alone, they are related with the structure of the Sun, as well. From here, we can say that the most important variation in the last SAC was seen for plage regions and this variation also effected the emergence of magnetic field from the Sun.

As a final result, small SGs (especially AB and C) were lower in cycle 23 than in cycle 22 while large SGs (DEF) were higher (or comparable with SAC 22) values in SAC 23. Large SGs were especially effective in the second maximum (after the year 2002) and FA were even higher from the beginning of 2001 opposite to Ca II K-flux and the covered area of plages on the Sun decreased after the middle of the year 2002. The FA were evaluated before the evolution of plage regions, so they would reached maximum phase before than the PA. In SAC 23, large SGs reached their maximum number about two years later than the small SGs, this difference was more efficient for magnetic field than TSI because magnetic field is related with small SGs while TSI values are related with large SGs. Large SGs follow SACs well-ordered than small SGs while small SGs move closer to the variation of ISSN through the SACs.

As we know, FA have a strong influence on the TSI, however, local magnetic field is strongly related with plage regions and the decrease in plage regions caused the decrease of the magnetic field intensity. For Ca II K-flux and plage variations, small SGs, especially C type SGs were important while FA were correlated with large SGs. FA and Ca II K-flux were separated from the synchronization in the beginning of 1999 which was the year where FA started to increase while Ca II K-flux was decreasing.

## 5. Conclusions

Normally, the ascending phases of SACs are characterized by a relatively higher activity than the descending ones, but the descending phase of SAC 23 and the early stage of the ascending phase of SAC 24 had shown a particularly high-level of activity. This anomalous activity from the beginning of SAC 23 can play a significant role in further development of SAC 24 and in the forecasting of parameters for the future cycles. Our current results show the importance of solar surface indices (especially plage regions) to the length of 11yr SAC and their relation between the time distribution of SGs depending on their types. In the last solar activity cycle (SAC 23), the covered surfaces by plages on the Sun were smaller in addition to the 13 years length of the cycle and it might have caused the increase of the ratio between facula and plage, and the magnetic flux from the Sun was in a lower values. The evolution and triple connection of faculae, Ca II K-flux and plage will answer our questions for the future SACs, so we need more observations of these parameters in the following cycles. In spite of a large amount of data is available for plage regions, the reliable detection of plage areas still need to be put beyond doubt.

In our paper, we compared the relation between solar surface indices with the number and the different type of sunspots/SGs, and found an important connection between them. Namely, large SGs follow SACs well-ordered than the small SGs, and C type and AB type SGs follow this coherence in the second and third orders. The uncertainty in the recent index values depending on the 13 years length of SAC 23 is already known, however, we relate these unexpected changes in solar surface indices with the size and complexity of different type of SGs for the first time. This is the key innovation from the previous studies brought by our approach. In SAC 23, large SGs reached their maximum number about 2 years later than the small SGs, this difference was more distinctive for magnetic field than TSI because magnetic field relates with the plage regions. In the same way, the FA which are connected with TSI were evaluated before the evolution of plage regions, so they would reach maximum phase before than the PA. In this respect, we found a linear relation between large (DEF type) SGs and FA, and mostly C type (and secondly DEF type) SGs and PA. Ca II K-flux variations are also comparable with C and DEF type SGs.

As we know, sunspots on the solar surface layers are connected with the physical variations of inner parts of the Sun. In the last SAC 23, the number of large SGs were higher or it was comparable with the previous SACs in spite of the 13 years length of SAC 23 and this shows the complex structure inside of the Sun in SAC 23 than the SACs 21 and 22. The dynamo theorem and/or helio-seismic oscillations inside the Sun will be proved our findings about the relation between FA, Ca II K-flux and PA and the different size of SGs in a good way. However, our first inventions changed the general definitions about the SACs and we first applied the different types of SGs to the solar indices which are related with the physical changes occur inside the Sun. We realized from these results that the inner parts of the Sun remarkably changed in SAC 23 and we plan to compare our current results with one of the theorems (dynamo or helio-seismic oscillations) above.

**Acknowledgements**
The authors would like to thank to the members of National Geographical Data Center (NGDC) and World Data Center (WDC) for making USAF_MWL (mainly Rome), and Learmonth proxies accessible for obtaining and further processing; Solar Irradiance Platform (SOLAR2000) historical irradiances are provided courtesy of W. Kent Tobiska and Space Environment Technologies. These historical irradiances have been developed with partial funding from the NASA UARS, TIMED, and SOHO missions; Wilcox Solar Observatory data used in this study was obtained via the web site http://wso.stanford.edu courtesy of J.T. Hoeksema for magnetic field data and the Wilcox Solar Observatory is currently supported by NASA; and the newest data of Stanford Observatory (SFO), California State University, Northridge is used courtesy of Dr. Angela M. Cookson. The new Sunspot Number data used in this article are produced and distributed by the World Data Center SILSO hosted that the Royal Observatory of Belgium, in Brussels. We thank the Director of the WDC-SILSO, Prof. Dr. Frédéric Clette, for providing us useful guidance and advices for the preparation of this article, and also to Boğaziçi University (Scientific Research Project No: 8563) and The Scientific and Technological Research council of Turkey (TÜBITAK) for their financial supports.

The authors also would like to thank anonymous referee(s) for very useful comments and guidance that improved the presentation of the paper.